\documentclass[pra, 12pt, showpacs, showkeys, eqsecnum, nofootinbib]{revtex4}

\begin{document}
\title{Symmetries in open quantum dynamics}
\author{Thomas F. Jordan}
\email[email: ]{tjordan@d.umn.edu}
\affiliation{Physics Department, University of Minnesota, Duluth, Minnesota 55812} 

\begin{abstract}
Simple examples are used to introduce and examine symmetries of open quantum dynamics that can be described by unitary operators. For the Hamiltonian dynamics of an entire closed system, the symmetry takes the expected form which, when the Hamiltonian has a lower bound, says that the unitary symmetry operator commutes with the Hamiltonian operator. There are many more symmetries that are only for the open dynamics of a subsystem. Examples show how these symmetries alone can reveal properties of the dynamics and reduce what needs to be done to work out the dynamics. A symmetry of the open dynamics of a subsystem can even imply properties of the dynamics for the entire system that are not implied by the symmetries of the dynamics of the entire system. The symmetries are generally not related to constants of the motion for the open dynamics of the subsystem. There are many symmetries that cannot be seen in the Schr\"{o}dinger picture as symmetries of dynamical maps of density matrices for the subsystem.

There are symmetries of the open dynamics of a subsystem that depend only on the dynamics. In the simplest examples, these are also symmetries of the dynamics of the entire system. There are many more symmetries, of a new kind, that also depend on correlations, or absence of correlations, between the subsystem and the rest of the entire system, or on the state of the rest of the entire system.
\end{abstract}


\maketitle

\newpage

\section{Introduction}\label{one}

There are symmetries of open quantum dynamics, described by unitary symmetry operators, that are not symmetries of the Hamiltonian dynamics of the entire closed system. We use simple examples to examine their properties. Our definition of symmetry is stated in terms of physically meaningful numbers. When it is applied to the Hamiltonian dynamics of an entire closed system, it takes the expected form which, when the Hamiltonian has a lower bound, says that the unitary symmetry operator commutes with the Hamiltonian operator. There are many more symmetries for the open dynamics of a subsystem. Examples show that these symmetries alone can reveal properties of the dynamics and reduce what needs to be done to work out the dynamics. A symmetry of the open dynamics of a subsystem can even imply properties of the dynamics for the entire system that are not implied by the symmetries of the dynamics of the entire system.

The open dynamics of a subsystem is described by a completely positive map when the initial density matrix for the entire system is a product of a density matrix for the subsystem and a density matrix for the rest of the entire system. Then our definition of a symmetry described by a unitary symmetry operator for the subsystem is just that the change of the density matrix for the subsystem is the same whether the unitary symmetry operator is applied before or after the map. Our definition also applies when the open dynamics of the subsystem is not described by a completely positive map, and it allows unitary symmetry operators that are not just for the subsystem.

Symmetries can look different for open quantum dynamics than for the complete quantum dynamics of a closed system. Structures where we might see symmetries are changed. The Schr\"{o}dinger picture is quite different.\cite{Nielsen2000,Breuer2002,AttalH,AttalM,AttalR,Alicki2007,Rivas2011} There may be no Schr\"{o}dinger equation, no wave function or state vector, because the state of the subsystem can be a mixed state described by a density matrix, not a pure state described by a state vector, even when the the state of the entire system that contains the subsystem is a pure state described by a state vector and the dynamics of the entire system is described by a Schr\"{o}dinger equation. The state of the subsystem can change between more or less pure or mixed as a pure state of the entire system changes in time. The time dependence of the density matrix for the subsystem may be described by a Gorini-Kossakowski-Sudarshan/Lindblad equation \cite{GKS,Lindblad} when necessary assumptions are satisfied or approximations are made. Completely positive maps of density matrices \cite{Kraus1971,Kraus1983} may be used when the density matrix for the entire system is a product of the density matrix for the subsystem and a density matrix for the rest of the entire system and, since that condition changes in time, maps with various different properties \cite{pechukas94,alicki95,pechukas95,stelmachovic01a,jordan04,jordan05,jordan05b,jordan08,Rybar,jordan09,jordan10} also may be brought in, further developing the basic picture \cite{ECGMathewsRau,jordan1961map,jordan1962map} of quantum dynamics as linear maps of density matrices.

The open dynamics of a subsystem does not look so different in the Heisenberg picture. The operators that represent the physical quantities of the subsystem are changed in time the same as for any other physical quantities of the entire system. The dynamics for the subsystem is seen simply by looking only at changes for the physical quantities of the subsystem. We will take this point of view here to look at symmetries of open quantum dynamics that can be described by unitary symmetry operators. This lets us use a framework that is the same for the symmetries of the dynamics of the entire system. It helps us see which symmetries come in only with the open dynamics for the subsystem, not as symmetries of the dynamics of the entire system. It also helps us see which symmetries depend only on the dynamics and which depend also on correlations, or absence of correlations, between the subsystem and the rest of the entire system or on the state of the rest of the entire system. It gives us exact equations and lets us avoid concerns about approximations used to write equations of motion for density matrices in the Schr\"{o}dinger picture.

In the quantum mechanics of an entire closed system, a unitary operator $U$ describes a symmetry for the quantum dynamics generated by a Hamiltonian operator $H$ if $U$ commutes with $H$. In terms of mean values, this means that
\begin{equation}
\label{mvbasic}
\text{Tr} \left[We^{itH}U^{\dagger}QUe^{-itH}\right] = \text{Tr} \left[WU^{\dagger}e^{itH}Qe^{-itH}U\right]
\end{equation}
for density matrices $W$ for states and operators $Q$ for physical quantities, for any time $t$. In Section II, we will see that, conversely, if Eq.(\ref{mvbasic}) holds for all $W$ and $Q$, and the spectrum of $H$ has a lower bound, then $U$ commutes with $H$. We will see that a physically equivalent conclusion about the symmetry of the dynamics generated by $H$ is obtained without the assumption that the spectrum of $H$ has a lower bound. The mean values in Eq.(\ref{mvbasic}) are physically meaningful numbers. They include the mean values for projection operators, which are the probabilities the states assign to possible values of physical quantities. The statement about mean values made with Eq.(\ref{mvbasic}) describes the symmetry in physical terms.

\newpage

This is for a quantum system that is closed, which means there is no need to consider that it might interact with anything else. An open quantum system is a subsystem $S$ of a larger system and interacts with the subsystem $R$ that is the remainder, or rest of the larger system (which could be a reservoir). We will consider the open dynamics for $S$ that is the result of the dynamics generated by a  Hamiltonian operator $H$ in the entire system of $S$ and $R$ combined. The mean values $\text{Tr} [WQ]$ for the operators $Q$ for the physical quantities of $S$ are changed to $\text{Tr} [We^{itH}Qe^{-itH}]$. Our definition of symmetry for open quantum dynamics is that a unitary operator $U$ describes a symmetry for the open dynamics of $S$ if Eq.(\ref{mvbasic}) holds just for the operators $Q$ for the physical quantities of $S$. We assume it holds for all the states of $S$. And the states of $R$? Correlations between $S$ and $R$? We will consider different possibilities.

In Section II, we assume that Eq.(\ref{mvbasic}) holds for all the states of the entire system of $S$ and $R$. The symmetries of the open dynamics of $S$ apply to all the states of $S$ and do not depend on the states of $R$ or on correlations, or absence of correlations, between $S$ and $R$. We call these \textit{independent} symmetries. We find that, at least in simple examples, this often implies that $U$ commutes with $H$; then there are no independent symmetries for the open dynamics of $S$  beyond those that are symmetries for the entire dynamics of $S$ and $R$.

In Section IV, we assume at first that there are no correlations between the states of $S$ and $R$ and we assume that Eq.(\ref{mvbasic}) holds for all the states of $S$ but only for particular states of $R$. Then, to complete Section IV, we admit correlations between $S$ and $R$ and assume that Eq.(\ref{mvbasic}) holds for all the states of $S$ but only for particular states of $R$ and particular correlations between the states of $S$ and $R$. In that section, the symmetries of the open dynamics of $S$ apply to all the states of $S$ but do depend on the state of $R$ and on correlations, or absence of correlations, between $S$ and $R$. We call these \textit{dependent} symmetries. This is a new kind of symmetry, different from that of the dynamics of an entire closed system. We consider only a few examples. Further results from collaboration are being reported separately.\cite{Seo}

The symmetries are generally not related to constants of the motion for the open dynamics of the subsystem. This is discussed in Section III.

We give most of our attention to symmetries described by unitary operators $U$ that are just for $S$; they do not involve $R$. In general, symmetries can be described by unitary operators $U$ that involve both $S$ and $R$. In particular, there are symmetries described by unitary operators that are just for $R$. They do not involve $S$. They come from outside $S$. They can show how the open dynamics of $S$ may depend on the state of $R$. There are examples in Sections II.B.2 and IV.A.

We follow common physics practice and write a product of operators for separate systems, for example a product of Pauli matrices $\Sigma$ and $\Xi$ for the two qubits considered in Section II.A, simply as $\Sigma \Xi$, not $\Sigma \otimes \Xi$. Occasionally we insert a $\otimes$ for emphasis or clarity. 

\section{Independent symmetries}\label{two}

Here we consider symmetries of the open dynamics of $S$ that do not depend on the states of $R$ or on correlations, or absence of correlations, between $S$ and $R$. Throughout this section, we assume that Eq.(\ref{mvbasic}) holds for all $W$ for all the states of the entire system of $S$ and $R$ combined, and for all the $Q$ for the physical quantities of $S$, which means that
\begin{equation}
\label{opbasic}
e^{itH}U^{\dagger}QUe^{-itH} = U^{\dagger}e^{itH}Qe^{-itH}U
\end{equation}
for all the $Q$ for $S$, and for any time $t$. This is a form our definition of symmetry takes when we work with independent symmetries. It says that the overall changes of the operators $Q$ for $S$ are the same whether the symmetry transformation is before or after the dynamics. Multiplying both sides of this Eq.(\ref{opbasic}) on the left by $U$ and on the right by $U^{\dagger}$ gives
\begin{eqnarray}
\label{opbasic2}
e^{itH}Qe^{-itH} & = & Ue^{itH}U^{\dagger}QUe^{-itH}U^{\dagger} \nonumber \\
& = & e^{itUHU^{\dagger}}Qe^{-itUHU^{\dagger}}.
\end{eqnarray}
This is another form our definition of symmetry takes when we work with independent symmetries. It says that the changes in time of the operators $Q$ for $S$ are the same for the dynamics generated by $UHU^{\dagger}$ as for the dynamics generated by $H$. 

The changes in time of the operators $Q$ for $S$ may be different in the dynamics generated by $U^{\dagger}HU$; this is shown by the example in Section II.E. The dynamics generated by $U^{\dagger}HU$ does give the same changes in time of the operators $Q$ for $S$ as the dynamics generated by $H$ if $U$ is an element of a group of unitary operators that represents a group of independent symmetries of the open dynamics, because then the requirement that the inverse of each element of the group also is an element of the group means that Eq.(\ref{opbasic}) holds when $U$ is replaced by $U^{\dagger}$. This is the case in the analog here of the familiar situation where a one-parameter group of symmetries is represented by a one-parameter group of unitary operators constructed from an Hermitian generator. Suppose $U$ is $e^{-i\theta J}$ with $J$ an Hermitian operator and $\theta $ a real parameter, and suppose that Eq.(\ref{opbasic}) holds for all real $\theta $. Then for a given $\theta $, it holds also for $-\theta $, so it holds for the given $\theta $ when $J$ is replaced by $-J$, which means that Eq.(\ref{opbasic}) holds with $U = e^{-i\theta J}$ replaced by $U^{\dagger} = e^{i\theta J}$.

Suppose $U_1$ and $U_2$ are operators for $S$; they do not involve $R$. If $U_1$ and $U_2$ represent independent symmetries for the open dynamics generated by $H$, then $U_1U_2$ also does, because Eq.(\ref{opbasic}) for $U_1U_2$ is implied by its holding successively for $U_1$ and then $U_2$. Whether a set of independent symmetries represented by operators $U$ for $S$ generates a group depends on whether the $U^{\dagger}$ operators represent independent symmetries for $H$.

The symmetries of the complete dynamics of an entire quantum system are simple in ways that the symmetries of the open dynamics of a subsystem are not. If $U$ describes a symmetry for the dynamics of an entire system, then Eq.(\ref{mvbasic}) and, equivalently,
\begin{equation}
\label{mvbasicS}
\text{Tr} \left[Ue^{-itH}We^{itH}U^{\dagger}Q\right] = \text{Tr} \left[e^{-itH}UWU^{\dagger}e^{itH}Q\right]
\end{equation}
hold for all $Q$ and $W$ for the entire system, so Eq.(\ref{opbasic}) holds for all the $Q$ for the physical quantities of the entire system and
\begin{equation}
\label{opbasicS}
Ue^{-itH}We^{itH}U^{\dagger} = e^{-itH}UWU^{\dagger}e^{itH}
\end{equation}
holds for all the density matrices $W$ for the states of the entire system. This shows that if $U$ describes a symmetry for an entire system, then $U^{\dagger}$ also does, whether $U$ commutes with $H$ or not, because the density matrices $W$ are linear combinations of projection operators that represent physical quantities, so Eq.(\ref{opbasic}) holds when $Q$ is replaced by a density matrix, and the operators $Q$ for the physical quantities are linear combinations of projection operators that are density matrices, so Eq.(\ref{opbasicS}) holds when $W$ is replaced by any operator that represents a physical quantity. Combining this with the observation just made that if $U_1$ and $U_2$ represent symmetries, then their product $U_1U_2$ also does, we see that the operators $U$ that describe symmetries for the dynamics of an entire system form a group. This is not generally true for the open dynamics of a subsystem.

We have a Heisenberg picture of the symmetries in Eq.(\ref{opbasic}) and a Schr\"{o}dinger picture in Eq.(\ref{opbasicS}), for the dynamics of an entire system. The Schr\"{o}dinger picture is not so simple for the open dynamics of a subsystem. From Eq.(\ref{mvbasic}) or, equivalently, Eq.(\ref{mvbasicS}) holding for all $Q$ for $S$, we have
\begin{equation}
\label{mvopbasicS}
\text{Tr}_R \left[Ue^{-itH}We^{itH}U^{\dagger}\right] = \text{Tr}_R \left[e^{-itH}UWU^{\dagger}e^{itH}\right].
\end{equation}
Since the dynamics is for the entire system of $S$ and $R$ combined, a change in the density matrix for $S$ is generally obtained by calculating the change of the density matrix $W$ for the entire system and taking the trace for $R$ at the end. The density matrix for the entire system, not just the density matrix for $S$, is needed at the start. 

The only situation that gives our symmetry framework a place in the Schr\"{o}dinger picture is when a symmetry is described by a unitary operator $U_S$ that is just for $S$, so it does not involve $R$, and there are no initial correlations between $S$ and $R$, so the density matrix $W$ for the entire system is a product $\rho_S \rho_R$ of density matrices $\rho_S$ for $S$ and $\rho_R$ for $R$. Then the dynamics changes $\rho_S$ in time $t$ to 
\begin{equation}
\label{Satt}
\Phi (\rho_S) = \text{Tr}_R \left[e^{-itH}\rho_S \rho_Re^{itH}\right]
\end{equation}
with a map $\Phi $ that is completely positive. From Eq.(\ref{mvopbasicS}), for $U$ an operator $U_S$ that is only for $S$, we have
\begin{eqnarray}
\label{rhosym}
U_S \Phi (\rho_S)U_S^{\dagger} & = & \text{Tr}_R \left[U_Se^{-itH}\rho_S \rho_Re^{itH}U_S^{\dagger}\right] \nonumber \\
& = & \text{Tr}_R \left[ e^{-itH}U_S \rho_S U_S^{\dagger} \rho_R e^{itH}\right] \nonumber \\
& = & \Phi (U_S\rho_S U_S^{\dagger})
\end{eqnarray}
which expresses the symmetry in terms of the map of the density matrices for $S$. A different choice of $\rho_R$ in Eq.(\ref{Satt}) defines a different map $\Phi $.

Other kinds of maps \cite{pechukas94,alicki95,pechukas95,stelmachovic01a,jordan04,jordan05b,jordan10} can be used when there are initial correlations between $S$ and $R$.  Different maps are defined by different correlations as well as by different states of $R$. Generally each map applies to a limited domain, a particular set of density matrices for $S$. The correlations and the map domains both can be changed by unitary operators $U_S$ that are only for $S$. It becomes difficult to describe the symmetries in terms of the maps.

A symmetry of the open dynamics of $S$ can imply properties of the dynamics, for the entire system of $S$ and $R$, that are not implied by the symmetries of the dynamics of the entire system. This is shown by an example worked out in Section II.F.

We can give at least partial answers right away to some of the most simple immediate questions about the new symmetries. Examples will eventually fill out the picture so we can see the kinds of symmetries that can occur. As a tool for our first calculations, we let

\vspace{-0.4cm}
\begin{equation}
\label{R(t)}
R(t) = Ue^{-itH}U^{\dagger}e^{itH}.
\end{equation}
By multiplying both sides on the right by $e^{-itH}U$, we see that
\begin{equation}
\label{R(t)HU}
Ue^{-itH} = R(t)e^{-itH}U
\end{equation}
for all $t$. Conversely, this Eq.(\ref{R(t)HU}) for all $t$ implies that Eq.(\ref{opbasic}) holds for all $t$ and all $Q$ for $S$ if the $R(t)$ commute with all the $Q$ for $S$. By multiplying both sides of Eq.(\ref{opbasic}) on the left by $Ue^{-itH}$ and on the right by $U^{\dagger}e^{itH}$ we see that the $R(t)$ for all $t$ commute with all the $Q$ for $S$. \\ \\
\textit{Theorem 1}. The $R(t)$ for all real $t$ commute with $e^{isH}Qe^{-isH}$ for all the $Q$ for $S$ and all real $s$. \\
\textit{Proof}. From Eq.(\ref{R(t)HU}) we get
\begin{eqnarray}
\label{prf1}
R(s+t)e^{-i(s+t)H}U & = & Ue^{-i(s+t)H} = Ue^{-isH}e^{-itH} \nonumber \\ & = & R(s)e^{-isH}Ue^{-itH} = R(s)e^{-isH}R(t)e^{-itH}U \\
\label{prf2}
R(s+t) & = & R(s)e^{-isH}R(t)e^{isH}
\end{eqnarray}
and we see that, since all the $Q$ for $S$ commute with both $R(s+t)$ and $R(s)$, they must commute with $e^{-isH}R(t)e^{isH}$ which means that every $R(t)$ commutes with $e^{isH}Qe^{-isH}$ for every $s$ for all the $Q$ for $S$. This completes the proof of Theorem 1.
 \\ \\
This theorem implies that the $R(t)$ commute with $[Q,H]$, $[[Q,H],H]$ and all the successive commutators $[[[Q,H],H] ...,H]$.

The operators required to commute with the $R(t)$ as a result of Theorem 1 have to be worked out specifically case by case. In some cases, $H$ commutes with the $R(t)$. Then Eq.(\ref{prf2}) implies that
\begin{equation}
\label{Rst} 
R(s+t)=R(s)R(t). 
\end{equation}
If the state-vector space is finite-dimensional, this  brings us to the conclusion that $U$ commutes with $H$. \\ \\ 
\textit{Theorem 2}. For operators on a finite-dimensional space, if the $R(t)$ commute with $H$ then $U$ commutes with $H$. \\
\textit{Proof}. If the $R(t)$ commute with $H$, then Eq.(\ref{Rst}) holds and implies that $R(t)=e^{-itG}$ with $G$ a Hermitian operator that commutes with $H$, and Eq.(\ref{R(t)HU}) implies that
\begin{equation}
\label{UHU2}
Ue^{-itH}U^{\dagger} = e^{-itG}e^{-itH}
\end{equation}
\vspace{-0.7cm}
so
\begin{equation}
\label{H+G2}
UHU^{\dagger}  =  H + G.
\end{equation}
Since the spectrum of $UHU^{\dagger}$ is the same as the spectrum of $H$, this says that the spectrum of $H$ is the same as the spectrum of $H + G$. For operators on a finite-dimensional space, this implies that $G$ is zero. This completes the proof of Theorem 2. \\  \\
In some cases, $R(t)$ commutes with all the operators for the entire system of $S$ and $R$. Then $R(t)$ must be a multiple of the identity operator for all $t$. If the spectrum of $H$ has a lower bound, this brings us to the conclusion, again, that $U$ commutes with $H$. \\ \\
\textit{Theorem 3}. If $R(t)$ is a multiple of the identity operator for all $t$, and the spectrum of $H$ has a lower bound, then $U$ commutes with $H$. \\
\textit{Proof}. Eq.(\ref{Rst}) implies that $R(t)=e^{-itr}$ with $r$ a real number, and then Eq.(\ref{R(t)HU}) implies that
\begin{equation}
\label{UHU}
Ue^{-itH}U^{\dagger} = e^{-itr}e^{-itH}
\end{equation}
\vspace{-0.7cm}
so
\begin{equation}
\label{H+r}
UHU^{\dagger}  =  H + r.
\end{equation}
Since the spectrum of $UHU^{\dagger}$ is the same as the spectrum of $H$, this says that the spectrum of $H$ is the same as the spectrum of $H + r$, which implies that $r$ is zero if the spectrum of $H$ has a lower bound. This completes the proof of Theorem 3. \\  \\
An example worked out in Section II.G shows that the assumption that the $R(t)$ are multiples of the identity operator is necessary for this theorem. An example worked out in Section II.E shows that the assumption that the spectrum of $H$ has a lower bound also is necessary for this theorem.

Now we can prove the statements that were left unproved in the Introduction. They are based on the assumption that Eq.(\ref{mvbasic}) holds for all $W$ and all $Q$ for the entire system of $S$ and $R$. This implies that Eq.(\ref{opbasic}) holds for all the $Q$ for the entire system of $S$ and $R$. Then the $R(t)$ commute with all the $Q$ for the entire system of $S$ and $R$, so the $R(t)$ must be multiples of the identity operator, and Theorem 3 implies that $U$ commutes with $H$ if the spectrum of $H$ has a lower bound.

Without the assumption that the spectrum of $H$ has a lower bound, we still get
\begin{equation}
\label{rHU}
Ue^{-itH} = e^{-itr}e^{-itH}U
\end{equation}
from Eq.(\ref{UHU}). This is a statement about the symmetry of the dynamics generated by $H$ that is physically equivalent to the statement that $U$ commutes with $H$; the phase factor $e^{-itr}$ makes no difference. This does not say, as the statement that $U$ commutes with $H$ does, that the values of the quantity represented by $H$ are the same as the values of the quantity represented by $UHU^{\dagger}$.

Having completed the Introduction, we go back to independent symmetries of the open dynamics of $S$, and consider simple examples.

\subsection{One and one qubits example} \label{qubit}

Let $S$ be a qubit described by Pauli matrices $\Sigma_1$, $\Sigma_2$, $\Sigma_3$ and $R$ a qubit described by Pauli matrices $\Xi_1$, $\Xi_2$, $\Xi_3$. In this example, there are no independent symmetries of the open dynamics of $S$ that are not also symmetries of the entire dynamics of $S$ and $R$. To see this, we will work out the commutators $[\Sigma_j, H]$ and $[[\Sigma_j, H], H]$ for any $H$ and apply Theorems 1 and 2.

We write the Hamiltonian as
\begin{equation}
\label{HPauli}
H = \frac{1}{2}\sum_{j=1}^3 \alpha_j \Sigma_j + \frac{1}{2}\sum_{j=1}^3 \beta_j \Xi_j + \frac{1}{2}\sum_{j=1}^3 \gamma_j \Sigma_j \Xi_j 
\end{equation}
with real numbers $\alpha_j $, $\beta_j $, $\gamma_j $. Any Hamiltonian can be put in this form \cite{zhang03a} by rotations of the $\Sigma_j$ and $\Xi_k $ that change $\sum_{j=1}^3 \sum_{k=1}^3 \gamma_{jk}\Sigma_j \Xi_k \; $ to  $\;  \sum_{j=1}^3 \gamma_j \Sigma_j \Xi_j $.

The $R(t)$ commute with $\Sigma_1$, $\Sigma_2$, $\Sigma_3$. By calculating the three commutators $[\Sigma_j, H]$ and multiplying each by each of the two $\Sigma_k$ for $k \neq j$, we see that Theorem 1 implies that the $R(t)$ commute with
\begin{equation}
\gamma_2 \Sigma_3 \Xi_2 - \gamma_3 \Sigma_2 \Xi_3, \, \, \, \gamma_3 \Sigma_1 \Xi_3 - \gamma_1 \Sigma_3 \Xi_1, \, \, \, \gamma_1 \Sigma_2 \Xi_1 - \gamma_2 \Sigma_1 \Xi_2, \nonumber
\end{equation}
\vspace{-1.0cm}
\begin{eqnarray}
\label{qlist1}
& \gamma_1 \Xi_1 - i\gamma_3 \Sigma_2 \Xi_3, \, \, \, & \gamma_1 \Xi_1 + i\gamma_2 \Sigma_3 \Xi_2, \nonumber \\
& \gamma_2 \Xi_2 - i\gamma_1 \Sigma_3 \Xi_1, \, \, \, & \gamma_2 \Xi_2 + i\gamma_3 \Sigma_1 \Xi_3, \nonumber \\
& \gamma_3 \Xi_3 - i\gamma_2 \Sigma_1 \Xi_2, \, \, \, & \gamma_3 \Xi_3 + i\gamma_1 \Sigma_2 \Xi_1,
\end{eqnarray}
which implies that the $R(t)$ commute with $\gamma_1 \Xi_1$, $\gamma_2 \Xi_2$, $\gamma_3 \Xi_3$.

If any two of $\gamma_1 $, $\gamma_2 $, $\gamma_3 $ are not zero, the $R(t)$ must commute with $\Xi_1$, $\Xi_2$, $\Xi_3$ so, since they also commute with $\Sigma_1$, $\Sigma_2$, $\Sigma_3$, the $R(t)$ must be multiples of the identity operator and either Theorem 2 or Theorem 3 implies that $U$ commutes with $H$. 

If $\gamma_1 $, $\gamma_2 $, $\gamma_3 $ are all zero, there is no interaction between $S$ and $R$. Then the dynamics in $S$ is generated just by the Hamiltonian $\frac{1}{2}\sum_{j=1}^3 \alpha_j \Sigma_j$ for $S$ independently of the dynamics generated by $\frac{1}{2}\sum_{j=1}^3 \beta_j \Xi_j$ for $R$. We will not consider this case.

We choose a representative of the three cases where just one of $\gamma_1 $, $\gamma_2 $, $\gamma_3 $ is not zero and consider the case where $\gamma_1 $ and $\gamma_2 $ are zero and $\gamma_3 $ is not zero. Then the $R(t)$ commute with $\Xi_3 $. Since they also commute with $\Sigma_1$, $\Sigma_2$, $\Sigma_3$, they must be functions of $\Xi_3 $, so
\begin{equation}
\label{a0a3}
R(t) = a_0(t) + a_3(t)\Xi _3 
\end{equation}
with complex numbers $a_0(t)$ and $a_3(t)$ for each $t$. By calculating $[[\Sigma_1, H], H]$, for the $H$ with $\gamma_1 $ and $\gamma_2 $ zero, and multiplying by $\Sigma_2$, we see that Theorem 1 implies that each $R(t)$ commutes with $\beta _1 \Xi_2 - \beta _2 \Xi_1$. This implies that either the $a_3(t)$ in Eq.(\ref{a0a3}) is zero and $R(t)$ is a multiple of the identity operator, or $\beta_1 $ and $\beta_2 $ are zero. Either way, $R(t)$ commutes with $H$ for each $t$ and Theorem 2 implies that $U$ commutes with $H$. This completes the proof that in this example there are no independent symmetries of the open dynamics of $S$ that are not also symmetries of the entire dynamics of $S$ and $R$.

\subsection{One and two qubits examples} \label{qubit2}

A minimal expansion of the preceding example will make room for different results. Let $S$ remain a single qubit described by Pauli matrices $\Sigma_1$, $\Sigma_2$, $\Sigma_3$ but now let $R$ be two qubits described by Pauli matrices $\Xi_1$, $\Xi_2$, $\Xi_3$ and $\Pi_1$, $\Pi_2$, $\Pi_3$.

\subsubsection{No new symmetries} \label{none}

We consider two different example Hamiltonians. The first is
\begin{equation}
\label{Hq3}
H = \Sigma_3 \Xi_3 + \Sigma_3 \Pi_3. 
\end{equation}
The $R(t)$ commute with $\Sigma_1$, $\Sigma_2$, $\Sigma_3$. By calculating the commutator $[\Sigma_2, H]$ and multiplying by $\Sigma_1$, we see that Theorem 1 implies that the $R(t)$ commute with $\Xi_3 + \Pi_3$, so the $R(t)$ commute with $H$ and Theorem 2 implies that $U$ commutes with $H$.

\subsubsection{Many new symmetries} \label{many}

Still looking for different results, we consider another Hamiltonian,
\begin{equation}
\label{Hq32}
H = \frac{1}{2}[\Sigma_3 \Xi_3 + \Xi_3 \Pi_3]. 
\end{equation}
The two terms of $H$ commute and the second term commutes with $\Sigma_1$, $\Sigma_2$, $\Sigma_3$, so the first term alone gives the changes in time of $\Sigma_1$, $\Sigma_2$, $\Sigma_3$, which are that $\Sigma_3$ is not changed and
\begin{eqnarray}
  \label{td12}
e^{itH}\Sigma_1 e^{-itH} & = & \Sigma_1 \cos t - \Sigma_2 \Xi_3 \sin t \nonumber \\
e^{itH}\Sigma_2 e^{-itH} & = & \Sigma_2 \cos t + \Sigma_1 \Xi_3 \sin t. 
\end{eqnarray}
The dynamics of $S$ is the same if the $\Pi$ qubit is removed from $R$.

Let $U_S$ be a unitary operator just for $S$, so it does not involve $R$. If $U_S$ describes an independent symmetry for the open dynamics of $S$, it does so as well when $R$ is the single qubit described by $\Xi_1$, $\Xi_2$, $\Xi_3$. From Section II.A, we know this implies that $U_S$ commutes with the first term of $H$. Then, since it commutes with the second term too, $U_S$ commutes with $H$ and describes a symmetry for the entire dynamics of $S$ and $R$ combined.

Let $U_R$ be a unitary operator just for $R$, so it does not involve $S$. Then $U_R$ commutes with $\Sigma_1$, $\Sigma_2$, $\Sigma_3$. It cancels out of the left side of Eq.(\ref{opbasic}) and satisfies that equation if it does not change the $e^{itH}Qe^{-itH}$ on the right side, which means here that is does not change $\Sigma_3$ and does not change the operators on the right sides of Eqs.(\ref{td12}). Symmetries are described by all the unitary operators made from $\Xi_3$ and $\Pi_1$, $\Pi_2$, $\Pi_3$. Those involving $\Pi_1$ and $\Pi_2$ do not commute with $H$; they describe symmetries of the open dynamics of $S$ that are not symmetries of the entire dynamics of $S$ and $R$ combined. Altogether the symmetries described by unitary operators for $R$ provide substantial information about the dynamics of $S$; they imply that the results the dynamics gives for $e^{itH}\Sigma_1 e^{-itH}$, $e^{itH}\Sigma_2 e^{-itH}$, $e^{itH}\Sigma_3 e^{-itH}$ do not depend on $\Xi_1$, $\Xi_2$ or $\Pi_1$, $\Pi_2$, $\Pi_3$.

\subsection{One and many angular momenta example} \label{JKLM}

It can still happen that $U$ must commute with $H$ when $S$ and $R$ are both large systems and when the part of each that interacts with the other is small. Here is an example. Suppose there are operators $J_1$, $J_2$, $J_3$ for $S$ that have angular-momentum commutation relations
\begin{equation}
\label{JJcom}
[J_j, J_k] = i\sum_{l=1}^3\epsilon_{jkl}J_l \, \, \,  \text{for} \, \, \, j,k = 1, 2, 3.
\end{equation}
We do not assume that these operators involve all of $S$ or even a large part of $S$. We do assume that the state-vector space for $S$ is finite dimensional. Suppose $R$ has two sets of angular-momentum operators, $K_1$, $K_2$, $K_3$ and $L_1$, $L_2$, $L_3$, so each set has angular-momentum commutation relations the same as Eq.(\ref{JJcom}) for the $J$, and the $K$ commute with the $L$. We do not assume that the operators $K$ and $L$ describe all of $R$. We do assume that the state-vector space for $R$ is finite dimensional. We will see that the example can easily be extended by putting more angular-momentum operators in with the $K$ and $L$.

We work with the operators
\begin{equation}
\label{Jpm}
J_\pm  = \frac{1}{\sqrt{2}}(J_1 \pm  iJ_2),
\end{equation}
which have commutation relations
\begin{equation}
\label{JJpmcom}
[J_+, J_-] = J_3, \, \, \, [J_3, J_\pm ] = \pm J_\pm ,
\end{equation}
and with the same combinations and commutation relations for the $K$ and $L$. Let
\begin{equation}
\label{HJKL}
H = J_+K_- + J_-K_+ + K_+L_- + K_-L_+.
\end{equation}
The $R(t)$ commute with $J_1$, $J_2$, $J_3$ and all the other operators for $S$. From the commutators
\begin{equation}
\label{HcomJKL}
[J_\pm , H] = \pm J_3K_\pm ,  
\end{equation}
\begin{eqnarray}
\label{HHcomJKL}
[[J_\pm , H], H] & = & J_3J_\pm K_3 + J_3K_3L_\pm   \nonumber \\
  & = & \pm J_+K_-K_\pm  \mp J_-K_+K_\pm ,
\end{eqnarray} 
we see that Theorem 1 implies that the $R(t)$ commute with $K_+$, $K_-$, $K_3$ and $L_+$, $L_-$, $L_3$, so the $R(t)$ commute with $H$ and Theorem 2 implies that $U$ commutes with $H$.

The same result may be obtained when more angular-momentum operators are added to the chain with the $K$ and $L$. If $M_+$, $M_-$, $M_3$ are added and
\begin{eqnarray}
\label{HJKLM}
H & = & J_+K_- + J_-K_+ + K_+L_- + K_-L_+  \nonumber \\
  &  & + \, L_+M_- + L_-M_+ ,
\end{eqnarray}
then the commutators $[[[J_\pm , H], H], H]$ have terms $\pm J_3K_3L_3M_\pm $ and show that Theorem 1 implies that the $R(t)$ commute with $M_+$, $M_-$, $M_3$ as well as $K_+$, $K_-$, $K_3$ and $L_+$, $L_-$, $L_3$, so the $R(t)$ still commute with $H$ and again Theorem 2 implies that $U$ commutes with $H$.

\subsection{One and one oscillators example with lower bound} \label{oscillatorw}

Let $S$ be an oscillator described by raising and lowering operators $A$ and $A^{\dagger}$ and $R$ an oscillator described by raising and lowering operators $B$ and $B^{\dagger}$ so
\begin{equation}
\label{ABcom}
[A,A^{\dagger}] = 1, \, \, \, \, [B,B^{\dagger}] = 1,
\end{equation}
and $A$ and $A^{\dagger}$ commute with $B$ and $B^{\dagger}$. The space of state vectors for $S$ and $R$ combined has orthonormal basis vectors $|m,n\rangle $ for $m=0,1,2,...$ and $n=0,1,2,...$ where
\begin{eqnarray}
\label{AmBn}
A|m,n\rangle  & = & m^\frac{1}{2} |m-1,n \rangle , \nonumber \\
A^{\dagger} |m,n\rangle  & = & (m+1)^\frac{1}{2} |m+1,n \rangle , \nonumber \\
B|m,n\rangle & = & n^\frac{1}{2} |m,n-1 \rangle ,\nonumber \\
B^{\dagger} |m,n\rangle  & = & (n+1)^\frac{1}{2} |m,n+1 \rangle .
\end{eqnarray}
Let
\vspace{-0.7cm}
\begin{eqnarray}
\label{How}
H & = & A^{\dagger}A + B^{\dagger}B + AB^{\dagger} + A^{\dagger}B \nonumber \\ & = & (A + B)^{\dagger}(A + B).
\end{eqnarray}

The $R(t)$ commute with $A$ and $A^{\dagger}$. From the commutators
\begin{equation}
\label{comos}
[A,H] = A + B, \, \quad \, [H,A^{\dagger} ] = A^{\dagger} + B^{\dagger} ,
\end{equation}
we see that Theorem 1 implies that the $R(t)$ commute with $B$ and $B^{\dagger}$. Then the $R(t)$ must be multiples of the identity operator and, since the spectrum of $H$ does have a lower bound, Theorem 3 implies that $U$ commutes with $H$. In this example, there are no independent symmetries of the open dynamics of $S$ that are not also symmetries of the entire dynamics of $S$ and $R$.

\subsection{One and one oscillators example without lower bound} \label{oscillatorwo}

If the terms without interactions are removed from the Hamiltonian, the spectrum of the Hamiltonian loses its lower bound. We get an example that shows that the assumption that the spectrum of $H$ has a lower bound is necessary for Theorem 3.

Again, let $S$ be an oscillator described by $A$ and $A^{\dagger}$ and $R$ an oscillator described by $B$ and $B^{\dagger}$, with Eqs.(\ref{ABcom}) and (\ref{AmBn}), but now let
\begin{equation}
\label{Hoswo}
H = AB^{\dagger} + A^{\dagger}B. 
\end{equation}
From the commutators
\begin{equation}
\label{comos}
[A,H] = B, \, \quad \, [H,A^{\dagger} ] = B^{\dagger} ,
\end{equation}
we see that Theorem 1 implies that the $R(t)$ commute with $B$ and $B^{\dagger}$ and conclude that the $R(t)$ must be multiples of the identity operator, the same as in the preceding example.

Then the $R(t)$ satisfy Eq.(\ref{Rst}), so $R(t)=e^{-itr}$ with $r$ a real number, and Eq.(\ref{R(t)HU}) implies that
\begin{equation}
\label{H+r2}
UHU^{\dagger}  =  H + r.
\end{equation}

To find a $U$, let
\begin{equation}
\label{JK}
J = \frac{1}{\sqrt{2}}(A + B), \, \quad \, K = \frac{1}{\sqrt{2}}(A - B).
\end{equation}
Then $J$ and $J^{\dagger}$ commute with $K$ and $K^{\dagger}$, and
\begin{equation}
\label{JKcom}
[J,J^{\dagger}] = 1, \, \quad \, [K,K^{\dagger}] = 1,
\end{equation}
so $J$, $J^{\dagger}$ and $K$, $K^{\dagger}$ are oscillator raising and lowering operators like $A$, $A^{\dagger}$ and $B$, $B^{\dagger}$, and
\begin{equation}
\label{HJK}
H = J^{\dagger}J - K^{\dagger}K.
\end{equation}
There are orthonormal vectors $|j,k \rangle_{J,K} $ for $j=0,1,2,...$ and $k=0,1,2,...$ where
\begin{equation}
\label{JKbasis}
J^{\dagger}J|j,k \rangle_{J,K}  = j|j,k \rangle_{J,K} , \, \quad \, K^{\dagger}K|j,k \rangle_{J,K}  = k|j,k \rangle_{J,K} .
\end{equation}
The space spanned by the vectors $|j,k \rangle_{J,K} $ is the same as the space spanned by the vectors $|m,n\rangle $ of Eqs.(\ref{AmBn}) because $A$ and $B$ are linear combinations of $J$ and $K$, so all the operators $A$, $A^{\dagger}$, $B$, $B^{\dagger}$, $J$, $J^{\dagger}$, $K$, $K^{\dagger}$ are defined on both spaces, and neither space has a partial subspace that is invariant for all the operators.
The $|j,k \rangle_{J,K} $ are eigenvectors of $H$ with
\begin{equation}
\label{Heigen}
H|j,k \rangle_{J,K}  = (j - k)|j,k \rangle_{J,K} , 
\end{equation}
so the spectrum of $H$ is all the integers, and the $r$ in Eq.(\ref{H+r2}) must be an integer. Let
\begin{eqnarray}
\label{Vjk}
V^{\dagger}|j,k \rangle_{J,K}  & = & |j,k-1 \rangle_{J,K} \, \, \,  \text{for} \, \, \,  k>j \nonumber \\
V^{\dagger}|j,k \rangle_{J,K}  & = & |j+1,k \rangle_{J,K} \, \, \,  \text{for} \, \, \,  j\geq k.
\end{eqnarray}It gives
\begin{eqnarray}
\label{HV}
HV^{\dagger} & = & V^{\dagger}(H+1), \nonumber \\
VHV^{\dagger} & = & H+1,
\end{eqnarray}
and $U = V^r$ satisfies Eq.(\ref{H+r2}) for any positive integer $r$. 

This Eq.(\ref{H+r2}) shows that Eq.(\ref{opbasic2}) holds for all the $Q$ for $S$ and $R$ as well as for the $Q$ for $S$. Although $U$ does not commute with $H$, the symmetry described by $U$ holds for the entire system of $S$ and $R$ as well as for $S$.

This example shows that the assumption that the spectrum of $H$ has a lower bound is necessary for Theorem 3. The $R(t)$ are multiples of the identity operator, but the spectrum of $H$ does not have a lower bound, and $U$ does not commute with $H$.

\subsection{Two oscillators and anything example} \label{oscillator2any}

We can expand the examples of Sections II.D and II.E to show that a symmetry of the open dynamics of $S$ can imply properties of the dynamics, for the entire system of $S$ and $R$, that are not implied by the symmetries of the dynamics of the entire system. Let $S$ be two oscillators described by $A$, $A^{\dagger}$ and $B$, $B^{\dagger}$, with Eqs.(\ref{ABcom}) and (\ref{AmBn}), and let
\begin{eqnarray}
\label{HJKM}
H & = & (AB^{\dagger} + A^{\dagger}B)M \nonumber \\
  & = & (J^{\dagger}J - K^{\dagger}K)M
\end{eqnarray}
as in Eqs.(\ref{Hoswo}) and (\ref{JK})-(\ref{HJK}). We do not assume that R is any particular system. We only assume that $M$ is an Hermitian operator for $R$ that has a discrete spectrum of eigenvalues $m$ that label basis vectors in the space of state vectors for $R$. For the entire system of $S$ and $R$ combined, there are orthonormal basis vectors $|j,k,m \rangle $ for $j=0,1,2,...$ and $k=0,1,2,...$, similar to those of Eqs.(\ref{JKbasis}) and (\ref{Heigen}), now with $m$ ranging over the eigenvales of $M$, and
\begin{eqnarray}
\label{JKbasis}
J^{\dagger}J|j,k,m \rangle & = & j|j,k,m \rangle \nonumber \\ 
K^{\dagger}K|j,k,m \rangle & = & k|j,k,m \rangle \nonumber \\
M|j,k,m \rangle & = & m|j,k,m \rangle \nonumber \\
H|j,k,m \rangle & = & (j - k)m|j,k,m \rangle .
\end{eqnarray}
As in Eq.(\ref{Vjk}), let
\begin{eqnarray}
\label{Vjkm}
V^{\dagger}|j,k,m \rangle  & = & |j,k-1,m \rangle \, \, \,  \text{for} \, \, \,  k>j \nonumber \\
V^{\dagger}|j,k.m \rangle  & = & |j+1,k,m \rangle \, \, \,  \text{for} \, \, \,  j\geq k.
\end{eqnarray}
This is an operator for $S$; it does not depend on $R$. It gives
\begin{eqnarray}
\label{HVM}
HV^{\dagger} & = & V^{\dagger}(H+M), \nonumber \\
VHV^{\dagger} & = & H+M.
\end{eqnarray}
Since $M$ commutes with $H$ and with all the operators $Q$ for $S$, we have
\begin{eqnarray}
\label{Vbasic2}
Ve^{itH}V^{\dagger}QVe^{-itH}V^{\dagger} & = & e^{it(H+M)}Qe^{-it(H+M)} \nonumber \\
& = & e^{itH}Qe^{-itH}
\end{eqnarray}
so Eq.(\ref{opbasic}) holds when $U$ is $V$, or when $U$ is $V^r$ for any positive integer $r$.

This is a symmetry of the open dynamics of $S$ that is not a symmetry of the dynamics for the entire system of $S$ and $R$; the $V$ and $U$ here do not commute with $H$. Nevertheless, this symmetry implies a property of the dynamics, for the entire system of $S$ and $R$, that is not implied by the symmetries of the dynamics for the entire system that are described by operators for $S$. If $U$ is an operator for $S$ that does commute with $H$, then $U$ commutes with $J^{\dagger}J - K^{\dagger}K$ and with
\begin{equation}
\label{Hf}
H_f = f(J^{\dagger}J - K^{\dagger}K)M 
\end{equation}
for any function $f$ of $J^{\dagger}J - K^{\dagger}K$, so $U$ describes a symmetry for the dynamics generated by $H_f$ as well as for the dynamics generated by the original $H$ where $f(j-k)$ is $j-k$. Knowing all these symmetries for the dynamics of the entire system provides no knowledge of $f$. Knowing that the $V$ of Eq.(\ref{Vjkm}) describes a symmetry of the open dynamics of $S$ makes it clear that the open dynamics is generated by the original $H$ where $f(j-k)$ is $j-k$.

We can see this because, from Eq.(\ref{Vjkm}),
\begin{equation}
\label{HVf}
H_fV^{\dagger}|j,k,m \rangle  =  f(j-k+1)mV^{\dagger}|j,k,m \rangle   
\end{equation}
so
\begin{equation}
\label{HVf}
VH_fV^{\dagger}  =  f(J^{\dagger}J - K^{\dagger}K+1)M   
\end{equation}
and Eq.(\ref{opbasic2}) implies that if V describes an independent symmetry for the open dynamics of $S$, then
\begin{equation}
\label{QbracH}
[Q,H_f] = [Q,VH_fV^{\dagger}]   
\end{equation}
and
\begin{equation}
\label{Qbracf}
[Q, f(J^{\dagger}J - K^{\dagger}K+1) - f(J^{\dagger}J - K^{\dagger}K)] = 0  
\end{equation}
for all the $Q$ for $S$, so the difference between $f(J^{\dagger}J - K^{\dagger}K+1)$ and $f(J^{\dagger}J - K^{\dagger}K)$ is a multiple of the identity operator for $S$. This means that $f$ can be taken to have a constant slope. Multiplying $M$ by this constant and dividing $f$ by it changes the slope of $f$ to $1$ and gives 
\begin{equation}
\label{fwewant}
f(J^{\dagger}J - K^{\dagger}K) = J^{\dagger}J - K^{\dagger}K + C 
\end{equation}
with $C$ a constant, so $H_f$ differs from the original $H$ where $f(j-k)$ is $j-k$ only by the operator $CM$ which commutes with $H$ and with all the operators $Q$ for $S$ and does not change the open dynamics for $S$.

\subsection{One and two oscillators example} \label{oscillator12}

Expansion of the example of Section II.D to one oscillator for $S$ and two oscillators for $R$ will give an example of an independent symmetry of the open dynamics of $S$ that is not a symmetry of the entire dynamics of $S$ and $R$. It will show that the assumption that the $R(t)$ are multiples of the identity operator is necessary for Theorem 3.

Let $S$ be an oscillator described by raising and lowering operators $A$ and $A^{\dagger}$ as in Section II.D and let $R$ be two oscillators described by raising and lowering operators $B$ and $B^{\dagger}$ and $C$ and $C^{\dagger}$ similar to those in Section II.D. Let
\begin{equation}
\label{J}
J = \frac{1}{\sqrt{3}}(A + B + C), \nonumber 
\end{equation} 
\begin{equation}
\label{K}
K = \frac{1}{\sqrt{2}}(B - C), \nonumber  
\end{equation}
\begin{equation}
\label{L}
L = \frac{1}{\sqrt{6}}(2A - B - C), 
\end{equation}
\begin{eqnarray} 
H & = & A^{\dagger}A + B^{\dagger}B + C^{\dagger}C + 3/2 \nonumber \\
 & & + AB^{\dagger} + A^{\dagger}B + BC^{\dagger} + B^{\dagger}C + CA^{\dagger} + C^{\dagger}A \nonumber \\ 
 & = & 3J^{\dagger}J + 3/2.
\end{eqnarray}
Then $J$ and $J^{\dagger}$ commute with $K$ and $K^{\dagger}$ and with $L$ and $L^{\dagger}$, and $K$ and $K^{\dagger}$ commute with $L$ and $L^{\dagger}$, and
\begin{equation}
\label{JKcom}
[J,J^{\dagger}] = 1, \, \quad \, [K,K^{\dagger}] = 1, \, \quad \, [L,L^{\dagger}] = 1.
\end{equation}

The $R(t)$ commute with $A$ and $A^{\dagger}$. From the commutators
\begin{equation}
\label{comos2}
[A,H] = A + B + C, \, \quad \, [H,A^{\dagger} ] = A^{\dagger} + B^{\dagger} + C^{\dagger},
\end{equation}
we see that Theorem 1 implies that the $R(t)$ commute with $B + C$ and $B^{\dagger} + C^{\dagger}$, so the $R(t)$ commute with $J$, $J^{\dagger}$, $L$, $L^{\dagger}$ and $H$. Then the $R(t)$ must be functions of $K$ and $K^{\dagger}$. Since the $R(t)$ commute with $H$, they satisfy Eq.(\ref{Rst}), so $R(t)=e^{-itG}$ with $G$ a function of $K$ and $K^{\dagger}$. To find a $U$ that describes an independent symmetry, we work with the orthonormal basis vectors $|j, k, l \rangle $ for $j=0,1,2,...$,   $k=0,1,2,...$ and $l=0,1,2,...$ where
\begin{eqnarray}
\label{JjKkLl}
J|j,k,l\rangle  & = & j^\frac{1}{2} |j-1,k,l \rangle , \nonumber \\
J^{\dagger} |j,k,l\rangle  & = & (j+1)^\frac{1}{2} |j+1,k,l \rangle , \nonumber \\
K|j,k,l\rangle  & = & k^\frac{1}{2} |j,k-1,l \rangle , \nonumber \\
K^{\dagger} |j,k,l\rangle  & = & (k+1)^\frac{1}{2} |j,k+1,l \rangle , \nonumber \\
L|j,k,l\rangle  & = & l^\frac{1}{2} |j,k,l-1 \rangle , \nonumber \\
L^{\dagger} |j,k,l\rangle  & = & (l+1)^\frac{1}{2} |j,k,l+1 \rangle , \nonumber \\
J^{\dagger}J|j,k,l \rangle & = & j|j,k,l \rangle , \nonumber \\
 K^{\dagger}K|j,k,l \rangle & = & k|j,k,l \rangle , \nonumber \\
 L^{\dagger}L|j,k,l \rangle & = & l|j,k,l \rangle , \nonumber \\ 
 H|j,k,l \rangle & = & (3j + 3/2)|j,k,l \rangle .
\end{eqnarray}
One choice for $U$ is to let
\begin{eqnarray}
\label{JjKkLlGU}
G|j,k=0,l\rangle  & = & 3|j,k=0,l \rangle , \nonumber \\
G|j,k,l\rangle  & = & 0 \, \, \, \text{for}\, \, \, k \neq 0, \nonumber \\
U^{\dagger} |j,k=0,l\rangle  & = & |j+1,k=1,l \rangle , \nonumber \\
U^{\dagger} |j,k=1,l\rangle  & = & |j,k=0,l \rangle , \nonumber \\
U^{\dagger} |j=0,k,l\rangle  & = & |j=0,k-1,l \rangle \, \, \, \text{for}\, \, \,  k>1 , \nonumber \\
U^{\dagger} |j,k,l\rangle  & = & |j,k,l \rangle \, \, \, \text{for}\, \, \, j>0, \, \, k>1.
\end{eqnarray}
This gives
\begin{eqnarray}
\label{HG}
HU^{\dagger} & = & U^{\dagger}(H + G) \nonumber \\
UHU^{\dagger} & = & H + G \nonumber \\
Ue^{-itH}U^{\dagger} & = & e^{-itG}e^{-itH} \nonumber \\
Ue^{-itH} & = & e^{-itG}e^{-itH}U
\end{eqnarray}
which gives Eq.(\ref{R(t)HU}), which implies that Eq.(\ref{opbasic}) holds for all $t$ and all the $Q$ for $S$ because the $R(t)$ do commute with all the $Q$ for $S$.. The spectrum of $H$ has a lower bound, and $H$ commutes with $G$ and the $R(t)$, but $U$ does not commute with $H$ or $G$. This example shows that the assumption that the $R(t)$ are multiples of the identity operator is necessary for Theorem 3. The arbitrary and awkward character of this example shows that other choices of $U$ would work as well.

We know that the changes in time of the operators $Q$ for $S$ are the same for the dynamics generated by $UHU^{\dagger}$ as for the dynamics generated by $H$. And for $U^{\dagger}HU\, $? In this example we have
\begin{equation}
\label{UHU2}
U^{\dagger}HU = H - U^{\dagger}GU,
\end{equation}
\begin{eqnarray}
\label{jkUGU}
U^{\dagger} GU|j,k=1,l\rangle  & = & 3|j,k=1,l \rangle \, \, \, \text{for}\, \, \, j \neq 0, \nonumber \\
U^{\dagger} GU|j,k,l\rangle  & = & 0 \, \, \, \text{for}\, \, \, k \neq 1 \, \, \, \text{or}\, \, \, j = 0,
\end{eqnarray}
and $A = \sqrt{\frac{1}{3}}J + \sqrt{\frac{2}{3}}L$. We see that $U^{\dagger}GU$ is a function of $J^{\dagger}J$ and $K^{\dagger}K$ and commutes with $L$, and $L^{\dagger}$, and
\begin{equation}
\label{AcomUGU}
[A, \; U^{\dagger} GU]|j=1,k=1,l\rangle  = \sqrt{3}|j=0,k=1,l \rangle , \nonumber 
\end{equation}
\begin{equation}
\label{AdcomUGU}
[A^{\dagger}, \; U^{\dagger} GU]|j=0,k=1,l\rangle  = -\sqrt{3}|j=1,k=1,l \rangle , 
\end{equation}
so $[A, \; U^{\dagger} GU]$ and $[A^{\dagger}, \; U^{\dagger} GU]$ are not zero. The changes in time of the operators $A$ and $A^{\dagger}$ for $S$ are not the same in the dynamics generated by $U^{\dagger}HU$ as in the dynamics generated by $H$.

\section{Constants of the motion}\label{three}

If we think about constants of the motion the same way we think about independent symmetries, and consider a statement that an operator $Q$ for $S$ represents a quantity that is a constant of the motion for the open dynamics of $S$, we could say that the statement should hold for all possible initial states of $S$ and for any state of $R$ and any correlations, or absence of correlations, between $S$ and $R$. Just for the mean value to be constant we would have
\begin{equation}
\label{constant}
\text{Tr} \left[We^{itH}Qe^{-itH}\right] = \text{Tr} \left[ WQ\right]
\end{equation}
for density matrices $W$ for all the states of the entire system of $S$ and $R$ combined, which implies that
\begin{equation}
\label{constan}
e^{itH}Qe^{-itH} = Q.
\end{equation}
which implies that $Q$ commutes with $H$. We would say that $Q$ can represent a constant of the motion for the open dynamics of $S$ only if $Q$ represents a constant of the motion for the dynamics of the entire system of $S$ and $R$ combined. In particular, when $Q$ is a unitary operator, we would also say that this implies that $Q$ describes a symmetry for the dynamics of the entire system of $S$ and $R$ combined.

On the other hand, if we think about constants of the motion the same way we think about dependent symmetries, we could say that an operator $Q$ for $S$ represents a quantity that is a constant of the motion for the open dynamics of $S$ if it is constant for all possible initial states of $S$ but only for particular states of $R$ or correlations, or absence of correlations, between $S$ and $R$. We could say it is a \textit{dependent constant of the motion}. We will see an example in Section IV.A of an Hermitian operator that is a generator of a one-parameter group of unitary operators that describe dependent symmetries but does not represent a dependent constant of the motion. More examples are being considered.\cite{Seo}

\section{Dependent symmetries}\label{four}

Now we consider dependent symmetries. At first, we assume there are no correlations between $S$ and $R$ and consider symmetries of the open dynamics of $S$ that depend on the state of $R$. We assume that the density matrix for $S$ and $R$ is a product $W=\rho_S\rho_R$ with $\rho_S$ a density matrix for $S$ and $\rho_R$ a density matrix for $R$. The mean value for a product of operators $A$ for $S$ and $B$ for $R$ is
\begin{equation}
\label{mvAB}
\langle AB \rangle  = \text{Tr} \left[WAB\right] = \text{Tr}_S \left[\rho_S A\right]\text{Tr}_R \left[\rho_R B\right] = \langle A \rangle \langle B \rangle .
\end{equation}
We assume that Eq.(\ref{mvbasic}) holds for all $\rho_S$ but only for particular $\rho_R$. The symmetries of the open dynamics of $S$ apply to all the states of $S$ but depend on the state of $R$. Then
\begin{equation}
\label{opbasicdep}
\text{Tr}_R \left[\rho_R e^{itH}U^{\dagger}QUe^{-itH}\right] = \text{Tr}_R \left[\rho_R U^{\dagger}e^{itH}Qe^{-itH}U\right]
\end{equation}
for all the $Q$ for $S$, and for any time $t$, but only for particular $\rho_R$. The changes for $S$ are the same whether the symmetry transformation is before or after the dynamics.

Dependent symmetries for unitary operators $U_S$ that are just for $S$, so they do not involve $R$, can be described in the Schr\"{o}dinger picture, using Eqs.(\ref{Satt}) and (\ref{rhosym}), when there are no initial correlations between $S$ and $R$. Different states of $R$ give different maps $\Phi $, and different maps $\Phi $ have different symmetries, so the symmetries depend on the states of $R$.

Suppose $U$ is an operator $U_S$ that is just for $S$; it does not involve $R$. Then multiplying both sides of Eq.(\ref{opbasicdep}) on the left by $U$ and on the right by $U^{\dagger}$ gives
\begin{eqnarray}
\label{opbasicdep2}
\text{Tr}_R \left[\rho_R e^{itH}Qe^{-itH}\right] & = & \text{Tr}_R \left[\rho_R U_Se^{itH}U_S^{\dagger}QU_Se^{-itH}U_S^{\dagger}\right] \nonumber \\
& = & \text{Tr}_R \left[\rho_R e^{itU_SHU_S^{\dagger}}Qe^{-itU_SHU_S^{\dagger}}\right].
\end{eqnarray}
The changes in time for $S$ are the same for the dynamics generated by $U_SHU_S^{\dagger}$ as for the dynamics generated by $H$. 

We know, from looking at independent symmetries, that generally the changes in time for $S$ may be different in the dynamics generated by $U^{\dagger}HU$ than in the dynamics generated by $H$. They are not different when $U$ is an operator $U_R$ that is just for $R$. Then $U$ and $U^{\dagger}$ cancel out of the left side of Eq.(\ref{opbasicdep}) and can be inserted on the right side to give
\begin{eqnarray}
\label{opbasicdep3}
\text{Tr}_R \left[\rho_R e^{itH}Qe^{-itH}\right] & = & \text{Tr}_R \left[\rho_R U_R^{\dagger}e^{itH}U_RQU_R^{\dagger}e^{-itH}U_R \right] \nonumber \\
& = & \text{Tr}_R \left[\rho_R e^{itU_R^{\dagger}HU_R}Qe^{-itU_R^{\dagger}HU_R}\right],
\end{eqnarray}
showing that the changes in time for $S$ are the same in the dynamics generated by $U_R^{\dagger}HU_R$ as in the dynamics generated by $H$. Also, when $U$ is an operator $U_R$ that is just for $R$, canceling $U$ and $U^{\dagger}$ out of the left side of Eq.(\ref{opbasicdep}) and moving $U$ to an equivalent position in the trace on the right side gives
\begin{equation}
\label{opbasicrho}
\text{Tr}_R \left[\rho_R e^{itH}Qe^{-itH}\right] = \text{Tr}_R \left[U_R\rho_R U_R^{\dagger}e^{itH}Qe^{-itH}\right].
\end{equation}
The changes in time for $S$ are the same for the state represented by $U_R\rho_R U_R^{\dagger}$ as for the state represented by $\rho_R $.

Suppose $U_1$ and $U_2$ are operators for $S$; they do not involve $R$. If $U_1$ and $U_2$ represent dependent symmetries for the open dynamics generated by $H$ and for the state of $R$ represented by $\rho_R $, then so does $U_1U_2$, because Eq.(\ref{opbasicdep}) for $U_1U_2$ is implied by its holding successively for $U_1$ and then $U_2$. Whether a set of dependent symmetries represented by operators $U$ for $S$ generates a group depends on whether the $U^{\dagger}$ operators represent dependent symmetries for the same $H$ and the same state of $R$.

Properties of the open dynamics often can be seen from a symmetry without working with the dynamics. Suppose $U$ is again an operator $U_S$ that is just for $S$; it does not involve $R$. If $U_S$ represents a dependent symmetry for the open dynamics generated by $H$ and for the state of $R$ represented by $\rho_R $, and if there are operators $Q$ and $Q_k$ for $S$ and numbers $d_k$ such that
\begin{equation}
\label{Qkcomb}
U_S^{\dagger}QU_S = \sum_k d_k Q_k,
\end{equation}
then Eq.(\ref{opbasicdep}) implies that
\begin{equation}
\label{Qkcombt}
U_S^{\dagger}\text{Tr}_R\left[\rho_R e^{itH}Qe^{-itH}\right]U_S = \sum_k d_k \text{Tr}_R \left[\rho_R e^{itH}Q_ke^{-itH}\right].
\end{equation}
In particular, if $Q$ commutes with $U_S$ then $\text{Tr}_R[\rho_R e^{itH}Qe^{-itH}]$ commutes with $U_S$; and if $Q$ anticommutes with $U_S$ then $\text{Tr}_R[\rho_R e^{itH}Qe^{-itH}]$ anticommutes with $U_S$. An example in Section IV.A shows how this can reduce what needs to be done to work out the dynamics.

The unitary symmetry operators, and the Hermitian operators that are generators for one-parameter groups of symmetry operators, generally do not represent constants of the motion for the open dynamics of $S$. This also is seen in the example of Section IV.A.

To consider symmetries that also depend on correlations between $S$ and $R$, we just assume that Eq.(\ref{mvbasic}) holds for density matrices $W$ that describe all the states of $S$ but only particular correlations between $S$ and $R$ and particular states of $R$. The changes for $S$ made by $U$ and the dynamics generated by $H$ are seen in the changes of the mean values of basic operators $Q$ for $S$ calculated with those $W$. This is illustrated in the example that follows.

\subsection{One and one qubits example} \label{qubitdep}

Let $S$ be a qubit described by Pauli matrices $\Sigma_1$, $\Sigma_2$, $\Sigma_3$ and $R$ a qubit described by Pauli matrices $\Xi_1$, $\Xi_2$, $\Xi_3$, as in Section II.A. Let
\begin{equation}
  \label{Hqd}
H = \frac{1}{2} \left[ \gamma_1 \Sigma_1 \Xi_1 + \gamma_2 \Sigma_2 \Xi_2 + \gamma_3 \Sigma_3 \Xi_3 \right].
\end{equation}
The three matrices $\Sigma_1 \Xi_1 $, $\Sigma_2 \Xi_2 $, $\Sigma_3 \Xi_3 $ commute with each other. (The different $\Sigma_j $ anticommute and the different $\Xi_j $ anticommute, so the different $\Sigma_j \Xi_j $ commute.) This allows us to easily compute
\begin{eqnarray}
  \label{US1}
e^{itH}\Sigma_1 e^{-itH} & = & \Sigma_1 e^{-it \gamma_2 \Sigma_2 \Xi_2 } e^{-it \gamma_3 \Sigma_3 \Xi_3 } \nonumber \\
 &=& \Sigma_1 \cos \gamma_2 t\cos \gamma_3 t + \underline{\Xi_1 \sin \gamma_2 t\sin \gamma_3 t} \nonumber \\
 && - \Sigma_2 \Xi_3 \cos \gamma_2 t\sin \gamma_3 t + \underline{\Sigma_3 \Xi_2 \sin \gamma_2 t\cos \gamma_3 t}
\end{eqnarray}
using the algebra of Pauli matrices, and similarly
\begin{eqnarray}
  \label{US2}
e^{itH}\Sigma_2 e^{-itH} &=& \Sigma_2 \cos \gamma_3 t\cos \gamma_1 t + \underline{\Xi_2 \sin \gamma_3 t\sin \gamma_1 t} \nonumber \\
 && - \underline{\Sigma_3 \Xi_1 \cos \gamma_3 t\sin \gamma_1 t} + \Sigma_1 \Xi_3 \sin \gamma_3 t\cos \gamma_1 t, 
\end{eqnarray}
\begin{eqnarray}
  \label{US3}
e^{itH}\Sigma_3 e^{-itH}  &=& \Sigma_3 \cos \gamma_1 t\cos \gamma_2 t + \Xi_3 \sin \gamma_1 t\sin \gamma_2 t \nonumber \\
 && - \underline{\Sigma_1 \Xi_2 \cos \gamma_1 t\sin \gamma_2 t} + \underline{\Sigma_2 \Xi_1 \sin \gamma_1 t\cos \gamma_2 t}. 
\end{eqnarray}

Let $U=\Sigma_3 $. Then Eq.(\ref{opbasicdep}) holds when $Q$ is $\Sigma_1 $, $\Sigma_2 $ or $\Sigma_3 $ if $\text{Tr}_R $ of each underlined term of Eqs.(\ref{US1})-(\ref{US3}) is zero, because $U$ and $U^{\dagger}$ cancel out of the left side of Eq.(\ref{opbasicdep}) after they change the sign of the whole left side when $Q$ is $\Sigma_1 $ or $\Sigma_2 $ and make no change when $Q$ is $\Sigma_3 $, and $U$ and $U^{\dagger}$ cancel out of the right side after they change the sign of each $\Sigma_1 $ and $\Sigma_2 $. The alternative Eq.(\ref{opbasicdep2}) also holds then, because changing $H$ to $UHU^{\dagger}$ just changes the signs of $\gamma_1$ and $\gamma_2$ and the underlined terms are the terms that change sign when this is done. Either way, we see that $U=\Sigma_3 $ describes a dependent symmetry of the open dynamics of $S$, when there are no correlations between the states of $S$ and $R$, if each underlined term is zero when $\Xi_1$, $\Xi_2$, $\Xi_3$ are replaced by $\langle \Xi_1 \rangle $, $\langle \Xi_2 \rangle $, $\langle \Xi_3 \rangle $.

There are various ways this can happen. Either
\begin{equation}
\label{ul1} 
\gamma_1  = 0 \, \, \,  \text{and} \, \, \, \gamma_2  = 0,
\end{equation}
\begin{equation}
\label{ul2} 
\text{or} \, \, \, \gamma_2  = 0, \, \, \gamma_3  = 0 \, \, \, \text{and} \, \, \, \langle \Xi_1 \rangle  = 0,
\end{equation}
\begin{equation}
\label{ul3}
\text{or} \, \, \, \gamma_3  = 0, \, \, \gamma_1  = 0 \, \, \, \text{and} \, \, \, \langle \Xi_2 \rangle  = 0,   
\end{equation}
\begin{equation}
\label{ul4}
\text{or} \, \, \, \langle \Xi_1 \rangle  = 0 \, \, \, \text{and} \, \, \, \langle \Xi_2 \rangle  = 0.   
\end{equation}
In the first case, $U=\Sigma_3 $ commutes with $H$. In the other three cases, the symmetry depends on the state of $R$. The quantity represented by $\Sigma_3 $ is a constant of the motion for the open dynamics of $S$ only in the first case where $\Sigma_3 $ commutes with $H$; we can see from Eq.(\ref{US3}) that $\langle \Sigma_3 \rangle $ changes in time for various states of $S$ if $\gamma_1 $ and $\gamma_2 $ are not both zero, so $\Sigma_3 $ can not represent a dependent constant of the motion for any state of $R$.

This symmetry alone implies that the underlined terms of Eqs.(\ref{US1})-(\ref{US3}) are zero because, according to Eqs.(\ref{Qkcomb})-(\ref{Qkcombt}), it requires $\text{Tr}_R[\rho_R e^{itH}\Sigma_1e^{-itH}]$ and $\text{Tr}_R[\rho_R e^{itH}\Sigma_2e^{-itH}]$ to anticommute with $\Sigma_3$ and requires $\text{Tr}_R[\rho_R e^{itH}\Sigma_3e^{-itH}]$ to commute with $\Sigma_3$. Stronger symmetry can imply more properties of the dynamics. The rotation operators
\begin{equation}
\label{rotR} 
U(u) = e^{-iu(1/2)\Sigma_3}
\end{equation}
for all real $u$ represent dependent symmetries, when there are no correlations between the states of $S$ and $R$, in the case where $\gamma_1$ and $\gamma_2$ are equal and $\langle \Xi_1 \rangle $ and $\langle \Xi_2 \rangle $ are zero; it is easy to check, using Eqs.(\ref{US1})-(\ref{US3}), that Eq.(\ref{opbasicdep}) is satisfied with $\Sigma_1$, $\Sigma_2$ or $\Sigma_3$ for $Q$. According to Eqs.(\ref{Qkcomb})-(\ref{Qkcombt}), this symmetry requires that $\text{Tr}_R[\rho_R e^{itH}\Sigma_1e^{-itH}]$ and $\text{Tr}_R[\rho_R e^{itH}\Sigma_2e^{-itH}]$ rotate like $\Sigma_1$ and $\Sigma_2$ when put between $U(u)^{\dagger}$ and $U(u)$, which implies that
\begin{eqnarray}
  \label{vec12rot}
\text{Tr}_R\left[\rho_R e^{itH}\Sigma_1e^{-itH}\right] & = & A_{11}(t)\Sigma_1 - A_{12}(t)\Sigma_2 \nonumber \\
\text{Tr}_R\left[\rho_R e^{itH}\Sigma_2e^{-itH}\right] & = & A_{11}(t)\Sigma_2 + A_{12}(t)\Sigma_1
\end{eqnarray}
with real functions $A_{11}(t)$ and $A_{12}(t)$. When this symmetry is assumed, only four of the twelve terms of Eqs.(\ref{US1})-(\ref{US3}) need to be calculated from the dynamics. The symmetry generator $\Sigma_3 $ can not represent a dependent constant of the motion for any state of $R$ when $\gamma_1$ and $\gamma_2$ are not zero, because then, again, Eq.(\ref{US3}) implies that $\langle \Sigma_3 \rangle $ changes in time for various states of $S$.

For any $\gamma_1, \gamma_2, \gamma_3$, when there are no correlations between the states of $S$ and $R$, there are one-parameter groups of dependent symmetries like the one, for example, described by the unitary operators $e^{-iu(1/2)\Xi_3}$ when $\langle \Xi_1 \rangle$ and $\langle \Xi_2 \rangle$ are zero. These unitary operators do not involve $S$. They commute with all the operators $Q$ made from $\Sigma_1$, $\Sigma_2$, $\Sigma_3$, so they cancel out of the left sidev of Eq.(\ref{opbasicdep}). They satisfy that equation because they also make no changes on the right side, which we can see by taking mean values on the right sides of Eqs.(\ref{US1}) - (\ref{US3}). Half the terms are not changed simply because they are zero. The form the results of the dynamics can take is restricted by both the symmetry and by this requirement of zero mean values for the state of $R$.

To admit correlations between $S$ and $R$, we ask whether Eq.(\ref{mvbasic}) holds when $Q$ is $\Sigma_1 $, $\Sigma_2 $, $\Sigma_3 $, for density matrices $W$ for all the states of $S$ but only for particular correlations 
\begin{equation}
\label{corrjk} 
\Gamma_{jk} = \langle \Sigma_j \Xi_k \rangle  - \langle \Sigma_j \rangle \langle \Xi_k \rangle \, \, \, \text{for}\, \, \, j,k,=1,2,3
\end{equation}
between $S$ and $R$ and particular states of $R$ described by $\langle \Xi_1 \rangle $, $\langle \Xi_2 \rangle $, $\langle \Xi_3 \rangle $. 
For $U=\Sigma_3 $ again, Eq.(\ref{mvbasic}) holds when $Q$ is $\Sigma_1 $, $\Sigma_2 $, $\Sigma_3 $ if the mean value of each underlined term of Eqs.(\ref{US1}) - (\ref{US3}) is zero. With correlations included, the things that have to be zero for this to happen are now either
\begin{equation}
\label{ul1cor} 
\gamma_1 \, \, \, \text{and} \, \, \, \gamma_2 ,
\end{equation}
\begin{equation}
\label{ul2} 
\text{or} \, \, \, \gamma_2 , \, \, \gamma_3 , \, \, \, \Gamma_{21} , \, \, \Gamma_{31} \, \, \text{and} \, \, \, \langle \Xi_1 \rangle ,
\end{equation}
\begin{equation}
\label{ul3}
\text{or} \, \, \, \gamma_3 , \, \, \gamma_1 , \, \, \, \Gamma_{32} , \, \, \Gamma_{12} \, \, \text{and} \, \, \, \langle \Xi_2 \rangle ,   
\end{equation}
\begin{equation}
\label{ul4}
\text{or} \, \, \, \Gamma_{12} , \, \, \Gamma_{21} , \, \, \Gamma_{31} , \, \, \Gamma_{32} , \, \, \langle \Xi_1 \rangle \, \, \text{and} \, \, \, \langle \Xi_2 \rangle .   
\end{equation}

\section{Outlook}\label{five}

There are many more symmetries for the open dynamics of a subsystem than for the complete dynamics of the closed system that contains it. We have seen this by looking at symmetries described by unitary operators. The unitary symmetry operators can be just for the subsystem $S$, or just for $R$, the rest of the closed system, or for the entire system of $S$ and $R$ combined. There are symmetries of a new kind that depend on correlations, or absence of correlations, between $S$ and $R$ or on the state of $R$. We have seen that the symmetries can reveal properties of the dynamics and reduce what needs to be done to work out the dynamics. A symmetry of the open dynamics of a subsystem can imply properties of the dynamics for the entire system that are not implied by the symmetries of the dynamics of the entire system.

These observations are a beginning. Further examples and applications should be explored with hope that some can be put to significant use. One step being reported separately is a collaboration looking at more examples of dependent symmetries.\cite{Seo}


\begin{thebibliography}{12}

\bibitem{Nielsen2000}M. A. Nielsen and I. L. Chuang, \textsl{Quantum Computation and Quantum Information} (Cambridge University Press, 2000).

\bibitem{Breuer2002}H. P. Breuer and F. Petruccione, \textsl{The Theory of Open Quantum Systems} (Oxford University Press, 2002).

\bibitem{AttalH}S. Attal, A. Joye, and C. A. Pillet (Eds.), \textsl{Open Quantum Systems I, The Hamiltonian Approach} (Springer Verlag, 2006).

\bibitem{AttalM}S. Attal, A. Joye, and C. A. Pillet (Eds.), \textsl{Open Quantum Systems II, The Markovian Approach} (Springer Verlag, 2006).

\bibitem{AttalR}S. Attal, A. Joye, and C. A. Pillet (Eds.), \textsl{Open Quantum Systems III, Recent Developments} (Springer Verlag, 2006).

\bibitem{Alicki2007}R. Alicki and K. Lendi, \textsl{Quantum Dynamical Semigroups and Applications} (Springer Verlag, 2007).

\bibitem{Rivas2011}A. Rivas and S. F. Huelga, \textsl{Open Quantum Systems: An Introduction} (Springer Verlag, 2011).

\bibitem{GKS}V. Gorini, A. Kossakowski, and E. C. G. Sudarshan,
``Completely positive dynamical semigroups of N-level systems,'' J. Math. Phys. {\bf 17}, 821-825 (1976).

\bibitem{Lindblad}G. Lindblad,
``On the Generators of Quantum Dynamical Semigroups,'' Commun. math. Phys. {\bf 48}, 119-130 (1976).

\bibitem{Kraus1971}K. Kraus,
``General state changes in quantum theory,'' Ann. Phys. (N. Y.) {\bf 64}, 311-335 (1971). 

\bibitem{Kraus1983}K. Kraus, \textsl{States, Effects and Operations: Fundamental Notions of Quantum Theory} (Springer Verlag, 1983).

\bibitem{pechukas94}P. Pechukas,
``Reduced dynamics need not be completely positive,'' Phys. Rev. Letters {\bf 73}, 1060-1062 (1994).

\bibitem{alicki95}R. Alicki,
``Comment on ``Reduced dynamics need not be completely positive'','' Phys. Rev. Letters {\bf 75}, 3020 (1995).

\bibitem{pechukas95}P. Pechukas,
``Pechukas replies to Comment on ``Reduced dynamics need not be completely positive'','' Phys. Rev. Letters {\bf 75}, 3021 (1995).

\bibitem{stelmachovic01a}P. Stelmachovic and V. Buzek,
``Dynamics of open quantum systems initially entangled with environment: Beyond the Kraus representation,'' Phys. Rev. A {\bf 64}, 062106 (2001).

\bibitem{jordan04}T. F. Jordan, A. Shaji, and E. C. G. Sudarshan,
``The dynamics of initially entangled open quantum systems,'' Phys. Rev. A {\bf 70}, 052110 (2004).

\bibitem{jordan05}T. F. Jordan,
``Affine maps of density matrices,'' Phys. Rev. A {\bf 71}, 034101 (2005).

\bibitem{jordan05b}T. F. Jordan, A. Shaji, and E. C. G. Sudarshan,
``Mapping the Schr{\"o}dinger picture of open quantum dynamics,'' Phys. Rev. A {\bf 73}, 012106 (2006).

\bibitem{jordan08}T. F. Jordan, A. Shaji, and E. C. G. Sudarshan,
``Markov approximations encounter map domains. A hazard of open quantum dynamics,'' Phys. Rev. A {\bf 77}, 032104 (2008).

\bibitem{Rybar}T. Rybar and M. Ziman,
``Repeatable quantum memory channels,'' Phys. Rev. A {\bf 78}, 052114 (2008).

\bibitem{jordan09}T. F. Jordan and A. Shaji,
``Repeatable procedures and maps in open quantum dynamics,'' Phys. Lett. A {\bf 373}, 4219 (2009).

\bibitem{jordan10}T. F. Jordan,
``Maps and inverse maps in open quantum dynamics,'' Ann. Phys. (N. Y.) {\bf 325}, 2075-2089 (2010).

\bibitem{ECGMathewsRau}E. C. G. Sudarshan, P. M. Mathews, and J. Rau,
``Stochastic dynamics of quantum-mechanical systems,'' Phys. Rev. {\bf 121}, 920-924 (1961).

\bibitem{jordan1961map}T. F. Jordan and E. C. G. Sudarshan,
``Dynamical mappings of density operators in quantum mechanics,'' J. Math. Phys. {\bf 2}, 772-775 (1961).

\bibitem{jordan1962map}T. F. Jordan, M. A. Pinsky, and E. C. G. Sudarshan,
``Dynamical mappings of density operators in quantum mechanics II. 
Time-dependent mappings,'' J. Math. Phys. {\bf 3}, 848-852 (1962).

\bibitem{Seo}T. F. Jordan and S. H. Seo, ``Symmetry examples in open quantum dynamics,'' arXiv:1408.4394.

\bibitem{zhang03a}J. Zhang, J. Vala, S. Sastry, and K. B. Whaley,
``Geometric theory of nonlocal two-qubit operations,'' Phys. Rev. A {\bf 67}, 042313 (2003).






\end{thebibliography}
\end{document}